\documentclass[twocolumn]{jpsj3} 
\title{First-Principles Dynamical Coherent-Potential Approximation 
Approach to the Ferromagnetism of Fe, Co, and Ni}

\author{Yoshiro \textsc{Kakehashi}\thanks{E-mail address:
yok@sci.u-ryukyu.ac.jp, to be published in J. Phys. Soc. Jpn.} and
M. Atiqur R. \textsc{Patoary}
}

\inst{
Department of Physics and Earth Sciences,
Faculty of Science, University of the Ryukyus, \\
1 Senbaru, Nishihara, Okinawa, 903-0213, Japan
}

\abst{
Magnetic properties of Fe, Co, and Ni at finite temperatures have been
investigated on the basis of the first-principles dynamical CPA
(Coherent Potential Approximation) combined with the LDA (Local Density
Approximation) + $U$ Hamiltonian in the Tight-Binding Linear Muffintin
Orbital (TB-LMTO) representation.  The Hamiltonian includes the
transverse spin fluctuation terms. Numerical calculations have been
performed within the harmonic approximation with 4th-order dynamical
corrections.  Calculated single-particle densities of states in the 
ferromagnetic state indicate that the dynamical effects reduce the 
exchange splitting, suppress the band width of the quasi-particle 
state, and causes incoherent excitations corresponding the 6 eV
satellites. Results of the magnetization vs temperature curves, 
paramagnetic spin susceptibilities, 
and the amplitudes of local moments are presented. 
Calculated Curie temperatures ($T_{\rm C}$) are reported to be 
1930K for Fe, 2550K for Co, and 620K for Ni; $T_{\rm C}$ for Fe and Co 
are overestimated by a factor of 1.8, while $T_{\rm C}$ in Ni agrees 
with the experimental result.  Effective Bohr magneton numbers
calculated from the inverse susceptibilities are 3.0 $\mu_{\rm B}$ (Fe), 
3.0 $\mu_{\rm B}$ (Co), and 1.6 $\mu_{\rm B}$ (Ni), being in
agreement with the experimental ones.  Overestimate of $T_{\rm C}$ in
Fe and Co is attributed to the neglects of the higher-order dynamical 
effects as well as
the magnetic short range order.  
}

\kword{
dynamical CPA, metallic magnetism, 
iron, cobalt, nickel, Curie temperature, effective
Bohr magneton number, excitation spectra
}

\begin{document}
\maketitle

\section{Introduction} 
The 3$d$ transition metals, Fe, Co, and Ni are well-known to show
the ferromagnetism with rather high Curie temperatures 1040 K, 1388 K, 
and 630 K~\cite{arrott67,colvin65,noakes66}. 
Magnetic properties of these materials are 
characterized by the itinerant as well as local-moment 
behaviors~\cite{kake04}.
According to the angle resolved photoemission 
spectroscopy~\cite{schaefer04,cui09}, 
quasiparticle $d$ bands with the Fermi surface are observed in these
transition metals, and
the noninteger ground-state magnetizations per atom in unit of
the Bohr magneton number ($\mu_{\rm B}$) are found~\cite{danan68,bes70}; 
2.2 $\mu_{\rm B}$ (Fe), 1.72 $\mu_{\rm B}$ (Co), and 0.62 $\mu_{\rm B}$ 
(Ni).  These results indicate that the 3$d$ electrons are itinerant.  
On the other hand, the Curie-Weiss law in the susceptibility with
atomic effective Bohr magneton number~\cite{fallot44}, 
the Brillouin-like magnetization
vs temperature curve~\cite{potter34,meyers51,weiss26}, 
and a large peak of the specific
heat at the Curie temperature ($T_{\rm C}$)~\cite{kraft66,brawn64,gonnelly71} 
are well explained by a simple local-moment model.  

The itinerant vs local-moment behavior in 3$d$ transition metals has 
been a long-standing problem in the metallic magnetism.
The Stoner model combined with the first-principles band theory yields
the Curie temperatures being much higher than the experimental ones;
6000K for Fe and 3000K for Ni~\cite{gunnarson77,staunton92}, 
and does not explain the Curie-Weiss law.
The theory has been much improved by taking into account spin
fluctuations.  Hubbard~\cite{hub79} and Hasegawa~\cite{hase79} 
proposed a single-site spin fluctuation theory (SSF) 
on the basis of the functional integral 
method~\cite{strat58,hub59,evan70,mora74}.  
The theory interpolates between the weak and strong
Coulomb interaction limits and explained qualitatively  the 
magnetization vs temperature curve as well as the Curie-Weiss law on 
the basis of the band model.
The theory however is based on the high-temperature approximation, 
{\it i.e.}, the static approximation which neglects the time
dependence of the field variables.  Therefore the SSF reduces to 
the Hartree-Fock approximation at the
ground state, and thus does not take into account electron correlations 
as found in the ground state 
theories~\cite{gutz63,gutz64,gutz65,hub63,hub64,kana63}.  

Kakehashi and Fulde~\cite{kake85}
proposed a variational theory which takes into account the electron
correlations at the ground state and reduces to the SSF at
high-temperatures.  They found that the reduction of the
Curie temperature by a factor of two in the case of Fe.  
Hasegawa~\cite{hase90}
developed the same type of theory on the basis of the Slave-Boson
functional integral approach.  Although these theories include the
ground-state electron correlations within the Gutzwiller-type 
approximation~\cite{gutz63,gutz64,gutz65,stoll78},
systematic improvement of thermodynamics is not easy in these approaches.  
We therefore proposed a theory called 
the dynamical coherent potential approximation 
(CPA)~\cite{kake92} 
which completely takes into account the dynamical charge and spin 
fluctuations within the single-site approximations, and clarified 
numerically the basic aspects of the theory with use of a 
Monte-Carlo technique.

In order to simplify numerical calculations, we proposed in the next
paper~\cite{kake02}, 
which we refer to I,  more analytic 
theory of the dynamical CPA using the harmonic approximation, 
and verified on the basis of the Hubbard model that the dynamical effects
reduce the Curie temperatures, cause the band narrowing for
quasiparticle states, and create the `6 eV' satellite peak in excitation
spectra. 
In our recent paper~\cite{kake08}, which we refer to II in the following, 
we proposed the first-principles dynamical CPA.  The theory combines the
dynamical CPA with the Local
Density Approximation (LDA) + U scheme~\cite{anis97} 
in the tight-binding 
linear muffin-tin orbital (TB-LMTO)~\cite{ander94} representation.  We 
investigated the dynamical effects on the magnetic properties of Fe 
and Ni within the 2nd-order dynamical corrections.
Quite recently, we have improved the first-principles dynamical CPA
taking into account the higher-order corrections~\cite{kake10}, 
and clarified the
systematic change of excitation spectra in 3$d$ transition metal 
series.

In this paper, we present numerical results of calculations for the
magnetic properties of Fe, Co, and Ni which are obtained by the
4th-order first-principles dynamical CPA.  We investigate the dynamical
effects on various quantities, and clarify the quantitative aspects of 
the first-principles dynamical CPA on the magnetic properties.

As we have proven in our previous papers~\cite{kake04,kake02-2}, 
the dynamical CPA is equivalent to the many-body CPA~\cite{hiro77} in
disordered alloys, the dynamical mean-field theory 
(DMFT)~\cite{mull89,jarr92,ohkawa92,geor93,geor96} 
in the metal-insulator transition,
and the projection operator CPA~\cite{kake04-1} 
in excitation problem in solids.  The first-principles DMFT calculations 
for Fe and Ni at the ground state have been
performed by Miura and Fujiwara~\cite{miura08} 
within the iterative perturbation method.
The finite-temperature DMFT calculations for Fe and Ni
have been performed by using the Hamiltonian without transverse spin
fluctuations~\cite{lich01}.  
We present here the finite-temperature results for the Hamiltonian with
transverse spin fluctuations including the results for Co.

In the following section, we outline the first-principles dynamical CPA
starting from the TB-LMTO type of Hamiltonian with intraatomic Coulomb 
interactions, and elucidate how to take into account the higher-order 
dynamical corrections using the asymptotic approximation.  
In \S 3, we present the numerical results of the
densities of states, effective potentials, magnetization vs temperature
curves, paramagnetic susceptibilities, and the amplitudes of local 
moments as a function of temperature.  
We will discuss on the dynamical effects on these physical
quantities and quantitative aspects of the theory.  
Especially, we will demonstrate that high-temperature physical
quantities such as the effective Bohr magneton number are qualitatively
described by the present theory.
We summarize our results in the last section \S 4, and discuss remaining
problems.

\section{First-principles TB-LMTO dynamical CPA}

We adopt in the present paper the TB-LMTO Hamiltonian combined with 
a LDA+U Coulomb interactions as follows~\cite{kake08}.  
\begin{eqnarray}
H = H_{0} + H_{1} ,
\label{hhat}
\end{eqnarray}
\begin{eqnarray}
H_{0} = \sum_{iL\sigma} (\epsilon^{0}_{L} - \mu) \, \hat{n}_{iL \sigma} 
+ \sum_{iL jL^{\prime} \sigma} t_{iL jL^{\prime}} \, 
a_{iL \sigma}^{\dagger} a_{jL^{\prime} \sigma} \ ,
\label{h0}
\end{eqnarray}
\begin{eqnarray}
H_{1}&=& \sum_{i} 
\Big[ \sum_{m} U_{0} \, \hat{n}_{ilm \uparrow} \hat{n}_{ilm \downarrow} 
+ {\sum_{m > m^{\prime}}} 
(U_{1}-\frac{1}{2}J) \nonumber \\
& &\hspace*{-5mm} \times \hat{n}_{ilm} \hat{n}_{ilm^{\prime}}
-{\sum_{m > m^{\prime}}} J   
\hat{\mbox{\boldmath$s$}}_{ilm} \cdot \hat{\mbox{\boldmath$s$}}_{ilm^{\prime}} 
\Big] \ . 
\label{h1}
\end{eqnarray}
Here we assumed a transition metal with an atom per unit cell.
$\epsilon^{0}_{L}$ is an atomic level on site $i$ and orbital $L$,
$\mu$ is the chemical potential, 
$t_{iL jL^{\prime}}$ is a transfer integral between orbitals $iL$ and 
$jL^{\prime}$. $L=(l,m)$ denotes $s$, $p$, and $d$ orbitals.
$a_{iL \sigma}^{\dagger}$ 
($a_{iL \sigma}$) is the creation (annihilation) operator for an
electron with orbital $L$ and spin $\sigma$ on site $i$, and 
$\hat{n}_{iL\sigma}=a_{iL \sigma}^{\dagger}a_{iL \sigma}$ is a charge
density operator for electrons with orbital $L$ and spin $\sigma$ on
site $i$. 
 
The Coulomb interaction term $H_{1}$ consists of the on-site
interactions between $d$ electrons ($l=2$).
$U_{0}$ ($U_{1}$) and $J$ denote the intra-orbital (inter-orbital)
Coulomb and exchange interactions, respectively.  
$\hat{n}_{ilm}$ ($\hat{\mbox{\boldmath$s$}}_{ilm}$) with $l=2$ is 
the charge (spin)
density operator for $d$ electrons on site $i$ and orbital $m$.
Note that the atomic level $\epsilon^{0}_{L}$ in 
$H_{0}$ is not identical with the LDA atomic level
$\epsilon_{L}$; $\epsilon^{0}_{L} = \epsilon_{L} - \partial
E^{U}_{\rm LDA}/\partial n_{iL\sigma}$. 
Here $n_{iL\sigma}$ is the charge density at the ground state, 
$E^{U}_{\rm LDA}$ is a LDA functional to the intraatomic Coulomb 
interactions~\cite{anis97,anis97-2}.

In the dynamical CPA, we transform in the free energy the interacting 
Hamiltonian $H_{1}$
into a one-body Hamiltonian with dynamical potential $v$ for 
time-dependent random
charge and exchange fields, using the
functional integral method~\cite{mora74,kake08}.  
Introducing a site-diagonal uniform medium, 
{\it i.e.}, a coherent potential $\Sigma$ into the potential part, 
we expand the correction $v-\Sigma$ with respect
to sites.  The zeroth term in the expansion is the free energy for 
a uniform medium, $\tilde{\cal F}[\Sigma]$.  
The next term is an impurity contribution to the
free energy.  The dynamical CPA neglects the higher-order terms
associated with inter-site correlations.
The free energy per atom is then given by~\cite{kake02,kake08}
\begin{eqnarray}
{\mathcal F}_{\rm CPA} = \tilde{\mathcal F}[\Sigma]
- \beta^{-1} {\rm ln} \, \int \Big[ \prod_{\alpha} 
\sqrt{\dfrac{\beta \tilde{J}_{\alpha}}{4\pi}}
d \xi_{\alpha} \Big] \,
{\rm e}^{\displaystyle -\beta E_{\rm eff}(\mbox{\boldmath$\xi$})} .
\nonumber\hspace*{-10mm}\\
\label{fcpa2}
\end{eqnarray}
Here $\beta$ is the inverse temperature,  
$\tilde{J}_{x}=\tilde{J}_{y}=\tilde{J}_{\bot}=[1-1/(2l+1)]J$, 
$\tilde{J}_{z} = U_{0}/(2l+1) + \tilde{J}_{\bot}$, and
$\mbox{\boldmath$\xi$} = (\xi_{x}, \xi_{y}, \xi_{z})$ is a static 
field variable on a site.  

The effective potential 
$E_{\rm eff}(\mbox{\boldmath$\xi$})$ in eq. (\ref{fcpa2}) consists of 
the static contribution $E_{\rm st}(\mbox{\boldmath$\xi$})$ and 
the dynamical correction term 
$E_{\rm dyn}(\mbox{\boldmath$\xi$})$ as follows.
\begin{eqnarray}
E_{\rm eff}(\mbox{\boldmath$\xi$}) = E_{\rm st}(\mbox{\boldmath$\xi$}) 
+ E_{\rm dyn}(\mbox{\boldmath$\xi$}) .
\label{eeff}
\end{eqnarray}
The former is given as
\begin{eqnarray}
E_{\rm st}(\boldsymbol{\xi}) &=& 
- \dfrac{1}{\beta} \sum_{mn} 
{\rm ln} \Big[ 
(1 \! - \! \delta v_{L\uparrow}(0)F_{L\uparrow}(i\omega_{n}))
(1 \! - \! \delta v_{L\downarrow}(0)\nonumber \\
& & \times F_{L\downarrow}(i\omega_{n}))
- \dfrac{1}{4} \tilde{J}^{2}_{\bot} \xi^{2}_{\bot} 
F_{L\uparrow}(i\omega_{n})F_{L\downarrow}(i\omega_{n})
\Big]     \nonumber \\
&  & 
+ \dfrac{1}{4} \Big[
- (U_{0}-2U_{1}+J) \sum_{m} \tilde{n}_{L}(\boldsymbol{\xi})^{2}\nonumber \\
& &- (2U_{1}-J) \tilde{n}_{l}(\boldsymbol{\xi})^{2}
+ \tilde{J}^{2}_{\bot} \xi^{2}_{\bot} + \tilde{J}^{2}_{z} \xi^{2}_{z}
\Big] .
\label{est2}
\end{eqnarray}
Here 
$\delta v_{L\sigma}(0) = v_{L\sigma}(0) - \Sigma_{L\sigma}(i\omega_{n})$, 
and $\xi^{2}_{\bot}= \xi^{2}_{x} + \xi^{2}_{y}$.  
$v_{L\sigma}(0)$ is a static potential given by 
$v_{L\sigma}(0) = [(U_{0}-2U_{1}+J)\tilde{n}_{lm}(\boldsymbol{\xi})+
(2U_{1}-J)\tilde{n}_{l}(\boldsymbol{\xi})]/2 -
\tilde{J}_{z}\xi_{z}\sigma/2$,
$\Sigma_{L\sigma}(i\omega_{n})$ is the coherent potential in
Matsubara frequency representation, and $\omega_{n}=(2n+1)\pi/\beta$.
The electron number $\tilde{n}_{L}(\boldsymbol{\xi})$ for a given 
$\boldsymbol{\xi}$ is expressed by means of an impurity Green function 
as
\begin{eqnarray}
\tilde{n}_{L}(\boldsymbol{\xi}) = \frac{1}{\beta} \sum_{n\sigma}
 G_{L\sigma}(\boldsymbol{\xi}, i\omega_{n}) ,
\label{nlxi}
\end{eqnarray}
and $\tilde{n}_{l}(\boldsymbol{\xi}) = 
\sum_{m} \tilde{n}_{L}(\boldsymbol{\xi})$.
The impurity Green function 
$G_{L\sigma}(\boldsymbol{\xi}, i\omega_{n})$ 
has to be determined self-consistently.  The explicit expression will 
be given later (see eq. (\ref{gimp})). 
 
The coherent Green function $F_{L\sigma}(i\omega_{n})$ in eq. (\ref{est2})
is defined by
\begin{eqnarray}
F_{L\sigma}(i\omega_{n}) = [(i\omega_{n} - \mbox{\boldmath$H$}_{0} 
- \mbox{\boldmath$\Sigma$}(i\omega_{n}))^{-1}]_{iL\sigma iL\sigma} .
\label{fls}
\end{eqnarray}
Here $(\mbox{\boldmath$H$}_{0})_{iL\sigma jL^{\prime}\sigma}$ 
is the one-electron Hamiltonian matrix for the noninteracting
Hamiltonian $H_{0}$, and 
$(\mbox{\boldmath$\Sigma$}(i\omega_{n}))_{iL\sigma jL^{\prime}\sigma} = 
\Sigma_{L\sigma}(i\omega_{n})\delta_{ij}\delta_{LL^{\prime}}$.

The dynamical potential $E_{\rm dyn}(\mbox{\boldmath$\xi$})$
in eq. (\ref{eeff}) has been obtained within the harmonic
approximation~\cite{kake02,kake08,amit71,dai91}. 
It is based on an expansion of $E_{\rm dyn}(\boldsymbol{\xi})$ with
respect to the frequency mode of the dynamical potential 
$v_{L\sigma\sigma^{\prime}}(i\omega_{\nu})$, where 
$\omega_{\nu}=2\nu\pi/\beta$.  The harmonic approximation is the neglect
of the mode-mode coupling terms in the expansion.  We have then  
\begin{eqnarray}
E_{\rm dyn}(\boldsymbol{\xi}) = - \beta^{-1} {\rm ln} 
\left[ 1 + \sum^{\infty}_{\nu=1} \,(\overline{D}_{\nu} -1) \right] .
\label{edyn1}
\end{eqnarray}
Here the determinant $D_{\nu}$ is a contribution from a dynamical
potential $v_{L\sigma\sigma^{\prime}}(i\omega_{\nu})$ with frequency
$\omega_{\nu}$, and the upper bar denotes a Gaussian
average with respect to the dynamical charge and exchange field
variables, $\zeta_{m}(i\omega_{n})$ and 
$\xi_{m\alpha}(i\omega_{n})$ ($\alpha = x, y, z$).  

The determinant $D_{\nu}$ is expressed as~\cite{kake08} 
\begin{eqnarray}
D_{\nu} = \prod_{k=0}^{\nu-1} \left[ \prod_{m=1}^{2l+1} D_{\nu}(k,m)
\right] ,
\label{dnu2}
\end{eqnarray}
%
%
%
%
%
\begin{eqnarray}
D_{\!\nu}(\!k,\!m\!) \!\!= \!\!\left| 
\begin{array}{@{\,}ccccccc@{\,}}
\ddots\!\!\!\!\!\!\!\!\!\!\!\!\! &         &         &      &    &   &  \\
       & 1       & 1           &        & 0  & &  \\
       & a_{\!-\!\nu\!+\!k}(\!\nu,\!m\!)\!\!\!\!\!\!\!\! & 1    & 1    &    & &  \\
       &                & a_{k}\!(\!\nu,\!m)\!\!\!\!\!\!\!\!\!\!\! & 1   & 1  & &  \\
       &            &      & a_{\nu\!+\!k}(\!\nu,\!m)\!\!\!\!\!\!\!\!\!\!\! & 1  & 1 & \\
       & 0       &      &       & a_{2\nu\!+\!k}(\!\nu,\!m)\!\!\!\!\!\!\!\! & & \\
       &          &      &          &                & \!\!\!\ddots \!\!\!\!\!\!\!\!\!\!\!\!& \\
\end{array}
\right| . \hspace{-2mm}
\label{dnukm}
\end{eqnarray}
Note that 1 in the above determinant denotes the $2 \times 2$ unit matrix, 
$a_{n}(\nu,m)$ is a $2 \times 2$ matrix
defined by 
\begin{eqnarray}
a_{n}(\nu,m)_{\sigma\sigma{\prime}} &=& 
\sum_{\sigma^{\prime\prime}\sigma^{\prime\prime\prime}
\sigma^{\prime\prime\prime\prime}} 
v_{L\sigma\sigma^{\prime\prime}}(i\omega_{\nu}) 
\tilde{g}_{L\sigma^{\prime\prime} \sigma^{\prime\prime\prime}}
(i\omega_{n}-i\omega_{\nu})\nonumber \\
& & \times v_{L\sigma^{\prime\prime\prime}\sigma^{\prime\prime\prime\prime}}
(-i\omega_{\nu}) 
\tilde{g}_{L\sigma^{\prime\prime\prime\prime} \sigma^{\prime}}(i\omega_{n}) \ ,
\label{annum}
\end{eqnarray}
\vspace{-4mm}
\begin{eqnarray}
v_{L\sigma\sigma^{\prime}}(i\omega_{\nu})&=& 
- \frac{1}{2}  \sum_{m^{\prime}} i A_{mm^{\prime}}
\zeta_{m^{\prime}}(i\omega_{\nu}) \delta_{l2}\delta_{\sigma\sigma^{\prime}} \nonumber \\
& &- \frac{1}{2} \sum_{\alpha} \sum_{m^{\prime}} 
B^{\alpha}_{mm^{\prime}} \xi_{m^{\prime}\alpha}(i\omega_{\nu})
\delta_{l2} 
(\sigma_{\alpha})_{\sigma\sigma^{\prime}} \ ,\nonumber\\
\label{dpot2}
\end{eqnarray}
\begin{eqnarray}
\tilde{g}_{L\sigma\sigma^{\prime}}(i\omega_{n}) = [(F_{L}(i\omega_{n})^{-1} - 
\delta v_{0})^{-1}]_{\sigma\sigma^{\prime}} \ .
\label{gst}
\end{eqnarray}
Here $\sigma_{\alpha}$ ($\alpha=x, y, z$) are the Pauli spin matrices.
$A_{mm^{\prime}} = U_{0}\delta_{mm^{\prime}} 
+ (2U_{1} - J)(1 - \delta_{mm^{\prime}})$, 
$B^{\alpha}_{mm^{\prime}} = J (1 - \delta_{mm^{\prime}})$ 
$(\alpha = x,y$), and $B^{z}_{mm^{\prime}} 
=  U_{0} \delta_{mm^{\prime}} + J (1 - \delta_{mm^{\prime}})$. 
$\tilde{g}_{L\sigma \sigma^{\prime}}(i\omega_{n})$ is the impurity
Green function in the static approximation,
$(F_{L}(i\omega_{n}))_{\sigma\sigma^{\prime}} = 
F_{L\sigma}(i\omega_{n})\delta_{\sigma\sigma^{\prime}}$, and 
$\delta v_{0}$ is defined by 
$(\delta v_{0})_{\sigma\sigma^{\prime}} = 
v_{L\sigma\sigma^{\prime}}(0) - 
\Sigma_{L\sigma}(i\omega_{n})\delta_{\sigma\sigma^{\prime}}$.

The determinant $D_{\nu}(k,m)$ is expanded with respect to the 
dynamical potential as follows. 
\begin{eqnarray}
D_{\nu}(k,m) = 1 + D^{(1)}_{\nu}(k,m) + D^{(2)}_{\nu}(k,m) + \cdots ,
\label{dnukm2}
\end{eqnarray}
\begin{eqnarray}
D^{(n)}_{\nu}(k,m) &=& \sum_{\alpha_{1}\gamma_{1} \cdots \alpha_{n}\gamma_{n}}
v_{\alpha_{1}}(\nu,m)v_{\gamma_{1}}(-\nu,m) \cdots \nonumber \\
& &\times v_{\alpha_{n}}(\nu,m)v_{\gamma_{n}}(-\nu,m) 
\hat{D}^{(n)}_{\{ \alpha\gamma \}}(\nu,k,m) \ .\nonumber \\
\label{dnnukm}
\end{eqnarray}
Here the subscripts 
$\alpha_{i}$ and $\gamma_{i}$ take 4 values $0$, $x$, $y$, and $z$,
and
\begin{eqnarray}
v_{0}(\nu,m) = - \dfrac{1}{2} i \sum_{m^{\prime}} A_{mm^{\prime}}
\zeta_{m^{\prime}}(i\omega_{\nu})\delta_{l2} \ ,
\label{v0num}
\end{eqnarray}
\begin{eqnarray}
v_{\alpha}(\nu,m) = - \dfrac{1}{2} \sum_{m^{\prime}} 
B^{\alpha}_{mm^{\prime}} \xi_{m^{\prime}\alpha}(i\omega_{\nu})\delta_{l2} \ , 
\hspace*{5mm} (\alpha=x,y,z) \ .\nonumber \hspace{-10mm}\\
\label{vanum}
\end{eqnarray}
Note that the subscript 
$\{ \alpha\gamma \}$ of $\hat{D}^{(n)}_{\{ \alpha\gamma \}}(\nu,k,m)$ 
in eq. (\ref{dnnukm}) denotes a set of 
$(\alpha_{1}\gamma_{1}, \cdots, \alpha_{n}\gamma_{n})$.

Substituting eq. (\ref{dnukm2}) into eq. (\ref{dnu2}) and taking the
Gaussian average, we reach
\begin{eqnarray}
E_{\rm dyn}(\boldsymbol{\xi}) = - \beta^{-1} {\rm ln} 
\left( 1 + \sum^{\infty}_{n=1} \sum^{\infty}_{\nu=1} 
\overline{D}_{\nu}^{(n)} \right) ,
\label{edyn2}
\end{eqnarray}
and
\begin{eqnarray}
\overline{D}^{(n)}_{\nu} &=& \dfrac{1}{(2\beta)^{n}} 
\sum_{\sum_{km} l(k,m)=n} \sum_{\{ \alpha_{j}(k,m)\} }
\sum_{\rm P}
\prod_{m=1}^{2l+1} \prod_{k=0}^{\nu-1}\nonumber \\
& &\hspace*{-3mm}\times\Bigg[ \Big( \prod_{j=1}^{l(k,m)} C^{\alpha_{j}(k,m)}_{mm_{\rm p}} \Big)
\hat{D}^{(l(k,m))}_{\{ \alpha\alpha_{{\rm p}^{-1}} \} }(\nu,k,m) \Bigg] .
\label{dnubarn}
\end{eqnarray}
Here each element of 
$\{ l(k,m)\}\, (k=0, \cdots, \nu-1, m=1, \cdots , 2l+1)$ has a value of 
zero or positive integer.
$\alpha_{j}(k,m)$ takes one of 4 cases $0$, $x$, $y$, and $z$.
$j$ denotes the $j$-th member
of the $(k,m)$ block with $l(k,m)$ elements. 
P denotes a permutation of a set $\{ (j,k,m) \}$; 
${\rm P} \{ (j,k,m) \} = \{ (j_{\rm p},k_{\rm p},m_{\rm p}) \}$.
$\alpha_{{\rm p}^{-1}}$ 
means a rearrangement of $\{ \alpha_{j}(k,m) \}$ according to 
the inverse permutation P${}^{-1}$.  
The coefficient $C^{\alpha}_{mm^{\prime}}$ in eq. (\ref{dnubarn}) 
is a Coulomb interaction defined by 
\begin{eqnarray}
C^{\alpha}_{mm^{\prime}} = \begin{cases}
-A_{mm^{\prime}} & (\alpha=0) \\
B^{\alpha}_{mm^{\prime}}  & (\alpha=x,y,z) \ .
\end{cases}
\label{cdef}
\end{eqnarray}
The frequency dependent factors 
$\hat{D}^{(n)}_{\{ \alpha\gamma \}}(\nu,k,m)$ in eq. (\ref{dnubarn})  
consist of a linear combination of $2n$ products of the static Green
functions.  Their first few terms have been given in Appendix A of 
our paper II~\cite{kake08}.

In the calculations of the higher-order dynamical corrections~\cite{kake10} 
$\hat{D}^{(n)}_{\{ \alpha\gamma \}}(\nu,k,m)$, 
we note that the coupling constants 
$B^{x}_{mm^{\prime}}=B^{y}_{mm^{\prime}}=J(1-\delta_{mm^{\prime}})$
are much smaller than $A_{mm^{\prime}}$ and $B^{z}_{mm^{\prime}}$
because $U_{0}$ and $U_{1} \gg J$.  Thus we neglect the transverse
potentials, $v_{x}(\nu, m)$ and $v_{y}(\nu, m)$. 
The approximation implies that $a_{n}(\nu, m)_{\sigma -\sigma}=0$.  
The determinant $D_{\nu}(k,m)$ in eq. (\ref{dnu2}) is then written 
by the products of the single-spin components as
\begin{eqnarray}
D_{\nu}(k,m) = D_{\nu\uparrow}(k,m)D_{\nu\downarrow}(k,m) .
\label{dnu3}
\end{eqnarray}
Here $D_{\nu\sigma}(k,m)$ is defined by eq. (\ref{dnukm}) in which 
the $2 \times 2$ unit matrices have been replaced by 1 ({\it i.e.}, 
$1 \times 1$ unit matrices), and the $2 \times 2$ matrices 
$a_{n}(\nu, m)$ have been replaced by the $1 \times 1$ matrices 
$a_{n}(\nu, m)_{\sigma\sigma}$.  The latter is given by
\begin{eqnarray}
a_{n}(\nu, m)_{\sigma\sigma} = \sum^{0,z}_{\alpha,\gamma}
 v_{\alpha}(\nu,m) v_{\gamma}(-\nu,m) \hat{h}_{\alpha\gamma\sigma}
e_{n\sigma}(\nu,m) ,
\label{annums}
\end{eqnarray}
\begin{eqnarray}
e_{n\sigma}(\nu,m) = \tilde{g}_{L\sigma}(n-\nu)\tilde{g}_{L\sigma}(n) .
\label{ennum}
\end{eqnarray}
Here $\hat{h}_{\alpha\gamma\sigma} = \delta_{\alpha\gamma} 
+ \sigma(1-\delta_{\alpha\gamma})$, and we used a notation 
$\tilde{g}_{L\sigma}(n) = \tilde{g}_{L\sigma\sigma}(i\omega_{n})$ for
simplicity. 
 
In order to reduce these summations, we make use of an
asymptotic approximation~\cite{kake02,kake10}.  
\begin{eqnarray}
e_{n\sigma}(\nu, m) \sim 
\overline{q}_{\nu} \, \big( 
\tilde{g}_{L\sigma}(n-\nu) - \tilde{g}_{L\sigma}(n) \big)
\ ,
\label{easym}
\end{eqnarray}
where $\overline{q}_{\nu}=\beta/2\pi\nu i$.
The approximation is justified in the
high-frequency limit where $\tilde{g}_{L\sigma}(n)$ written as
\begin{eqnarray}
\tilde{g}_{L\sigma}(n) = 
\dfrac{1}{i\omega_{n} - \epsilon^{0}_{L} + \mu - v_{L\sigma}(0)}
+ O\left(\dfrac{1}{(i\omega_{n})^{3}} \right) .
\label{gasym}
\end{eqnarray}

In the asymptotic approximation,
we obtain
\begin{eqnarray}
\hat{D}^{(n)}_{\{\alpha\gamma\}}(\nu, k, m) &=&
\sum^{n}_{l=0} \hat{D}^{(l)}_{\{\alpha_{1}\gamma_{1}
\cdots \alpha_{l}\gamma_{l}\}\uparrow}(\nu, k, m)\nonumber \\
& &\hspace*{-6mm}\times\hat{D}^{(n-l)}_{\{\alpha_{l+1}\gamma_{l+1}
\cdots \alpha_{n}\gamma_{n}\}\downarrow}(\nu, k, m) \ .
\label{dnag2}
\end{eqnarray}
Here we wrote the subscript at the r.h.s. explicitly to avoid
confusion.  Note that the values of $\alpha_{i}$ and $\gamma_{i}$ are
limited to $0$ or $z$ in the present approximation.
The spin-dependent quantities are given by~\cite{kake10}
\begin{eqnarray}
\hat{D}^{(l)}_{\{\alpha\gamma\}\sigma}(\nu, k, m) =
\Lambda^{(l)}_{\sigma}(\{\alpha\gamma\}) 
\frac{\overline{q}_{\nu}^{\,i}}{l\,!} B^{(l)}_{\sigma}(\nu, k, m)
\ ,
\label{dlags}
\end{eqnarray}
\begin{eqnarray}
\Lambda^{(l)}_{\sigma}(\{\alpha\gamma\}) = 
\begin{cases}
\ 1 & (\sigma=\uparrow) \\
(-1)^{l-n_{l}(\{\alpha\gamma\})} & (\sigma=\downarrow)
\end{cases}
\ ,
\label{lambda}
\end{eqnarray}
\begin{eqnarray}
B^{(l)}_{\sigma}(\nu, k, m) &=& 
\Big[ \prod^{l-1}_{j=0} \tilde{g}_{L\sigma}(j\nu+k) \Big] \nonumber \\
& &\hspace*{-3mm} + \sum^{l-1}_{i=0} \dfrac{(-)^{l-i} l!}{i! (l-i)!}
\Big[ \prod^{i-1}_{j=-(l-i)} \tilde{g}_{L\sigma}(j\nu+k) \Big]\nonumber \\
& &\hspace*{4mm}\times\Big[ 1 + \dfrac{l-i}{\overline{q}_{\nu}^{\,i}} 
\tilde{g}_{L\sigma}(i\nu+k) \Big] \ .
\label{bls}
\end{eqnarray}
Here $\hat{D}^{(0)}_{\{\alpha\gamma\}\sigma}(\nu, k, m) = 1$.
$n_{l}(\{\alpha\gamma\})$ is the number of $\{\alpha_{i}\gamma_{i}\}$ 
pairs such that $\alpha_{i}=\gamma_{i}$ among the $l$ pairs.
When there is no orbital degeneracy, eq. (\ref{bls}) reduces to the
result of the zeroth asymptotic approximation in our paper 
I~\cite{kake02}.

In the actual applications we make use of the exact form up to a certain
order of expansion in $\overline{D}^{(m)}_{\nu}$, 
and for higher order terms we adopt an approximate
form (\ref{dnag2}).  
In this way, we can take into account dynamical corrections
systematically starting from both sides, the weak interaction limit 
and the high-temperature one.

The coherent potential can be determined by the stationary condition
$\delta \mathcal{F}_{\rm CPA}/\delta \Sigma = 0$.  
This yields the CPA equation as~\cite{kake08} 
\begin{eqnarray}
\langle G_{L\sigma}(\mbox{\boldmath$\xi$}, i\omega_{n}) \rangle 
= F_{L\sigma}(i\omega_{n}) \ .
\label{dcpa3}
\end{eqnarray}
Here $\langle \ \rangle$ at the l.h.s. (left-hand-side) 
is a classical average taken with respect to the
effective potential $E_{\rm eff}(\mbox{\boldmath$\xi$})$.
The impurity Green function is given by 
\begin{eqnarray}
G_{L\sigma}(\mbox{\boldmath$\xi$}, i\omega_{l}) = 
\tilde{g}_{L\sigma\sigma}(i\omega_{l}) + 
\dfrac{\displaystyle \sum_{n} \sum_{\nu} 
\frac{\delta \overline{D}^{(n)}_{\nu}}
{\displaystyle \kappa_{L\sigma}(i\omega_{l})
\delta \Sigma_{L\sigma}(i\omega_{l})}}
{\displaystyle 1+ \sum_{n} \sum_{\nu} \overline{D}^{(n)}_{\nu}} \ .\nonumber \hspace*{-10mm}\\
\label{gimp}
\end{eqnarray}
Note that the first term at the r.h.s. (right-hand-side) is the impurity 
Green function in the static approximation, which is given 
by eq. (\ref{gst}). 
The second term is the dynamical corrections, and
$\kappa_{L\sigma}(i\omega_{l})= 1 - F_{L\sigma}(i\omega_{l})^{-2}
\delta F_{L\sigma}(i\omega_{l})/\delta \Sigma_{L\sigma}(i\omega_{l})$.

Solving the CPA equation (\ref{dcpa3}) self-consistently, we obtain 
the effective medium.  
The electron number on each orbital $L$ is then calculated from
\begin{eqnarray}
\langle \hat{n}_{L} \rangle = 
\dfrac{1}{\beta} \sum_{n\sigma} F_{L\sigma}(i\omega_{n}) \ .
\label{avnl}
\end{eqnarray}
The chemical potential $\mu$ is determined from the condition 
$n_{e} = \sum_{L} \langle \hat{n}_{L} \rangle$.
Here $n_{e}$ denotes the conduction electron number per atom.
The magnetic moment is given by
\begin{eqnarray}
\langle \hat{m}^{z}_{L} \rangle = 
\dfrac{1}{\beta} \sum_{n\sigma} \sigma F_{L\sigma}(i\omega_{n}) \ .
\label{avml}
\end{eqnarray}
In particular, the $l=2$ component of magnetic moment is expressed 
as
\begin{eqnarray}
\langle \hat{\boldsymbol{m}}_{l} \rangle = 
\langle \boldsymbol{\xi} \rangle \ . 
\label{avmd}
\end{eqnarray}
The above relation implies that the effective potential 
$E_{\rm eff}(\mbox{\boldmath$\xi$})$
is a potential energy for a local magnetic moment
$\mbox{\boldmath$\xi$}$.

In the numerical calculations, we took into account the dynamical
corrections up to the second order ($n \le 2$) exactly, 
and the higher-order terms
up to the fourth order within the asymptotic approximation.
Summation with respect to $\nu$ in Eqs. (\ref{edyn2})
and (\ref{gimp}) was taken up to $\nu = 100$ for $n=1$ and
$2$, and up to $\nu = 2$ for $n = 3, \ 4$.

When we solve the CPA equation (\ref{dcpa3}), we adopted a decoupling 
approximation to the thermal average of impurity Green 
function~\cite{kake81} for simplicity, 
{\it i.e.}, 
\begin{eqnarray}
\langle G_{L\sigma}(\xi_{z}, \xi^{2}_{\perp}, i\omega_{n}) \rangle && \nonumber \\
& &\hspace*{-38mm} = \sum_{q=\pm} \frac{1}{2}
\left( 1 + q \dfrac{\langle \xi_{z} \rangle}
{\sqrt{\langle \xi^{2}_{z} \rangle}} \right)
G_{L\sigma}(q\sqrt{\langle \xi^{2}_{z} \rangle}, 
\langle \xi^{2}_{\perp} \rangle, i\omega_{n}) \ .
\label{gapprox}
\end{eqnarray}
Here we wrote the static exchange field $\boldsymbol{\xi}$ as 
$(\xi_{z}, \xi^{2}_{\perp})$ so that the decoupling approximation we
made becomes clearer.
The approximation is correct up to the second moment ({\it i.e.},
$\langle \xi^{2}_{\alpha} \rangle$) and arrows us to 
describe the thermal spin fluctuations in a simpler way.

On the other hand, we adopted a diagonal approximation~\cite{kirk70} 
to the coherent Green function at the r.h.s. of eq. (\ref{dcpa3}). 
\begin{eqnarray}
F_{L\sigma}(n) = \int \dfrac{\rho^{\rm LDA}_{L}(\epsilon) d \epsilon}
{i\omega_{n} - \epsilon - \Sigma_{L\sigma}(i\omega_{n}) - 
\Delta\epsilon_{L}} \ .
\label{cohg2}
\end{eqnarray}
Here $\rho^{\rm LDA}_{L}(\epsilon)$ is the local density of states for the LDA
band calculation, and 
$\Delta \epsilon_{L} = (\epsilon_{L}-\epsilon^{0}_{L})\delta_{l2}$.
The approximation partly takes into account the effects of hybridization
between different $l$ blocks in the nonmagnetic state, but neglects the
effects via spin polarization.

The CPA equation (\ref{dcpa3}) 
with use of the decoupling approximation (\ref{gapprox})
yields an approximate solution to the full CPA equation.  
For the calculations of the single-particle densities of states, 
one needs more accurate solution for the CPA self-consistent equation.  
For this purpose, we adopted the following average $t$-matrix
approximation~\cite{korr58,ehren76} (ATA)  
after we solved eq. (\ref{dcpa3}) with the decoupling approximation 
(\ref{gapprox}).  
\begin{eqnarray}
 \Sigma^{\rm ATA}_{L\sigma}(i\omega_{n}) =  \Sigma_{L\sigma}(i\omega_{n}) + 
\dfrac{\langle G_{L\sigma}(\xi_{z}, \xi^{2}_{\perp}, i\omega_{n})
\rangle -F_{L\sigma}(i\omega_{n})}
{\langle G_{L\sigma}(\xi_{z}, \xi^{2}_{\perp}, i\omega_{n}) \rangle
F_{L\sigma}(i\omega_{n})} \ .
\label{ata}
\end{eqnarray}
Here the coherent potential in the decoupling
approximation is used at the r.h.s., but the full average 
$\langle \ \rangle$ of the 
impurity Green function is taken.  The ATA is a one-shot correction to
the full CPA.
%
%
\begin{figure}
\includegraphics[scale=0.75]{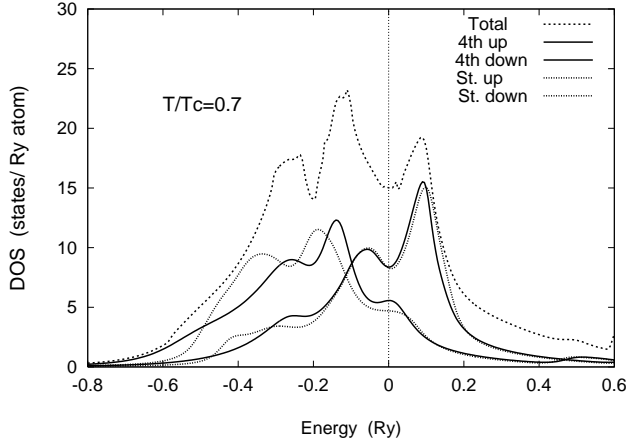}%
\caption{\label{figfedos}
Up and down $d$ partial densities of states (DOS) of Fe in the static 
approximation
 (dotted curves) and in the 4th-order dynamical CPA (solid curves).
The total DOS in the 4th-order dynamical CPA is shown
 by dashed curve.
}
\end{figure}
%
%
The coherent potential $\Sigma_{L\sigma}(z)$ on the real axis $z=\omega
+ i\delta$ is then calculated by using the Pad\'{e}
numerical analytic continuation method~\cite{vidberg77}. 
Here $\delta$ is an infinitesimal positive number.
The densities of states (DOS) as the single-particle excitations, 
$\rho_{L\sigma}(\omega)$ are 
calculated from the relation,
\begin{eqnarray} 
\rho_{L\sigma}(\omega) = - \frac{1}{\pi} \, {\rm Im} \, F_{L\sigma}(z) \ .
\label{dos}
\end{eqnarray}

\section{Numerical results}

In the numerical calculations, we adopted the lattice constants 
used by Andersen {\it et. al.}~\cite{ander94}, and 
performed the LDA calculations with use of the Barth-Hedin
exchange-correlation potential to make the TB-LMTO Hamiltonian (\ref{h0}).
For Fe and Ni, we adopted average Coulomb interaction parameters 
$\overline{U}$ and the average exchange interactions $\overline{J}$ 
used by Anisimov {\it et. al.}~\cite{anis97-2}, and for Co we adopted 
$\overline{U}$ obtained by Bandyopadhyay {\it et. al.}~\cite{bdyo89} 
and $\overline{J}$ obtained by the Hartree-Fock atomic
calculations~\cite{mann67} 
; $\overline{U}=0.169$ Ry and 
$\overline{J}=0.066$ Ry for bcc Fe, $\overline{U}=0.245$ Ry and 
$\overline{J}=0.069$ Ry for fcc Co, and $\overline{U}=0.221$ Ry and 
$\overline{J}=0.066$ Ry for fcc Ni. 
The intra-orbital Coulomb interaction $U_{0}$, inter-orbital Coulomb 
interaction $U_{1}$, 
and the exchange interaction energy parameter $J$ were calculated from 
$\overline{U}$ and $\overline{J}$ as 
$U_{0} = \overline{U} + 8\overline{J}/5$, 
$U_{1} = \overline{U}-2\overline{J}/5$, and $J=\overline{J}$, 
using the relation $U_{0} = U_{1}+2J$.  

Figure 1 shows the calculated densities of states (DOS) in the
ferromagnetic Fe.  Thermal spin fluctuations in the static
approximation broaden the band width for each spin band.
The DOS with the 4th-order dynamical charge and spin fluctuations shifts 
the main peak towards the Fermi level and causes a small hump around 
$\omega = -0.5$ Ry which corresponds to the `6 eV' satellite as found in
Ni at 6 eV below the Fermi level~\cite{himp79,east80,eber80,mar84,naka91}.
The present calculations with dynamical effects yields the ground-state
magnetization 2.58 $\mu_{\rm B}$, which is considerably larger than the
experimental value 2.22 $\mu_{\rm B}$.
The exchange splitting in the present calculations is therefore
somewhat overestimated.  When the exchange splitting is reduced, 
the main peak in the up spin band shifts up and the second
peak of the down spin band shifts down, so that one expects that 
the main peak around $\omega=-0.1$ Ry in the total DOS becomes
sharper and shifts towards the Fermi level.
%
%
\begin{figure}
\includegraphics[scale=0.85]{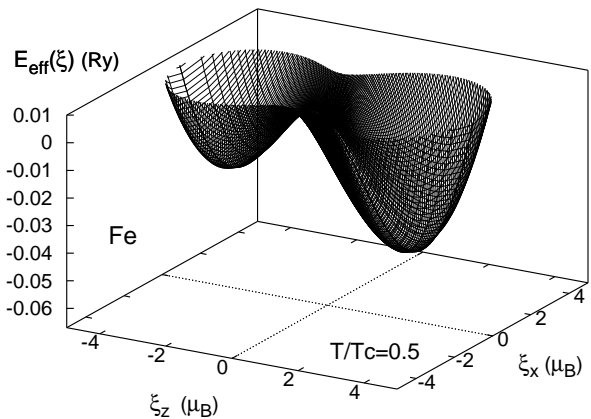}%
\caption{\label{figfeeff}
Effective potential for Fe at the temperature $T/T_{\rm C}=0.5$ on 
the $\xi_{x}-\xi_{z}$ plane.
}
\end{figure}
%
%

The effective potential in the dynamical CPA characterizes
spin fluctuations of the system.  
Unlike the single band model~\cite{kake02}, calculated
potential for ferromagnetic Fe has the double minimum structure as shown
in Fig. 2.  This implies that 
Fe local magnetic moments show large thermal spin fluctuations
which change magnetic moments in direction.  In order to clarify
the dynamical effects on the effective potential, we plotted the
dynamical contribution 
$E_{\rm dyn}(\boldsymbol{\xi}) = E_{\rm eff}(\boldsymbol{\xi}) - 
E_{\rm st}(\boldsymbol{\xi})$
in Fig. 3.
%
%
\begin{figure}
\includegraphics[scale=0.85]{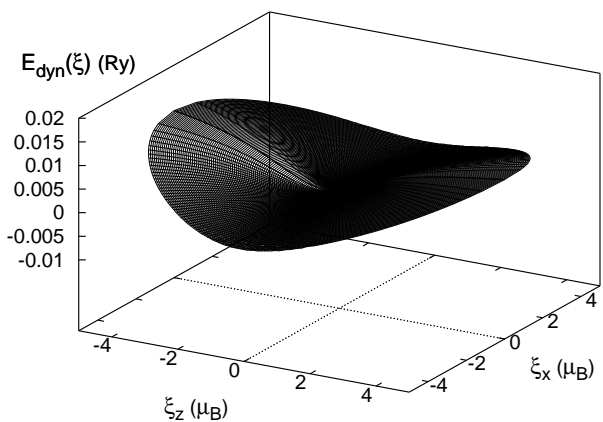}%
\caption{\label{figfedyn}
Dynamical contribution to the effective potential for Fe at 
$T/T_{\rm C}=0.5$.
}
\end{figure}
%
%
The dynamical potential of Fe has a saddle point at the origin; 
$E_{\rm dyn}(\boldsymbol{\xi})$ shows the minimum along the $z$ axis, and the
maximum along the $x$ ($y$) axis at the origin.  
It indicates that the dynamical
effects reduce the longitudinal amplitude of magnetic moments and
enhance the transverse spin fluctuations.  
These effects are enhanced with increasing temperatures.

The magnetization vs temperature curves of Fe in various approximations 
are presented in Fig. 4.
We obtained the Curie temperature $T_{\rm C}=2070$K in the static
approximation.  It is lower than that of the Hartree-Fock one
(12200K) by a factor of 6.  
The second-order dynamical corrections reduces $T_{\rm C}$
by 50K as shown in Fig. 4.  The 4th-order dynamical corrections further
reduce $T_{\rm C}$ by 90K as compared with the 2nd-order ones, and 
yield $T_{\rm C}=1930$K.  The result is
comparable to the value $T_{\rm}=1900$K based on the DMFT without
transverse spin fluctuations~\cite{lich01}, but 
still overestimates $T_{\rm C}$ by a factor of 1.8 as
compared with the experimental one (1040K)~\cite{arrott67}.
%
%
\begin{figure}
\includegraphics[scale=0.75]{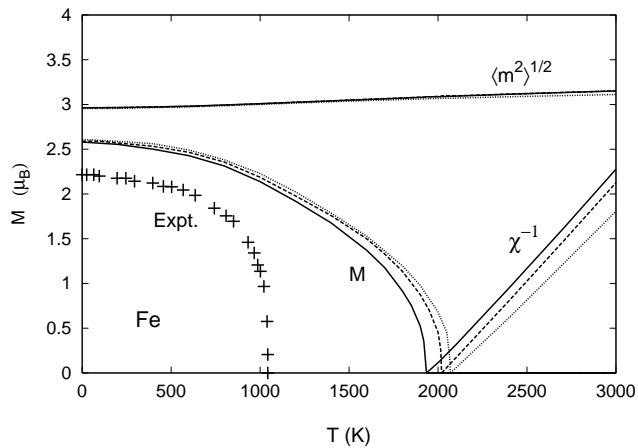}%
\caption{\label{figfemt}
Magnetization vs temperature curves ($M-T$), inverse susceptibilities 
($\chi^{-1}$), and the amplitudes of local magnetic moments 
($\langle \boldsymbol{m}^{2} \rangle^{1/2}$) for Fe in the static
 approximation (dotted curves), the 2nd-order dynamical CPA 
(dashed curves), and the 4th-order dynamical CPA (solid curves), 
respectively. Experimental $M-T$ curve~\cite{potter34} is shown by $+$ 
points. 
}
\end{figure}
%
%
Calculated inverse susceptibilities follows the Curie-Weiss law.  We
obtained the effective Bohr magneton number as 3.1
$\mu_{\rm B}$ (static approximation), 3.0 $\mu_{\rm B}$ (2nd-order
dynamical CPA), and 3.0 $\mu_{\rm B}$ (4th-order dynamical CPA),
respectively.  These values are in good agreement with the experimental
value~\cite{fallot44} 3.2 $\mu_{\rm B}$.  
The amplitudes of local magnetic moments 
$\langle \boldsymbol{m}^{2} \rangle^{1/2}$ show a weak temperature
dependence and take a value 3.1 $\mu_{\rm B}$ at 2000K irrespective of 
details of approximations as shown in Fig. 4.  
It should be noted that the
calculated effective Bohr magneton number approximately equals to the
amplitude of local moment; the Rhodes-Wohlfarth ratio is 1 in agreement
with the experimental fact.
%
%
\begin{figure}
\includegraphics[scale=0.75]{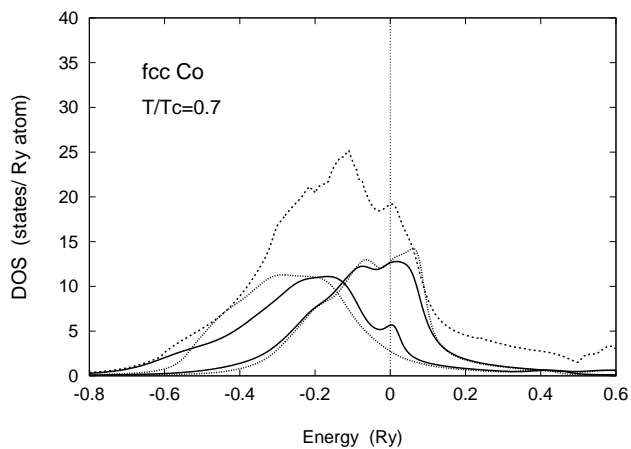}%
\caption{\label{figcodos}
Up and down $d$ partial DOS of Co in the static approximation
 (dotted curves) and in the 4th-order dynamical CPA (solid curves).
The total DOS in the 4th-order dynamical CPA is shown
 by dashed curve.
}
\end{figure}
%
%

In the case of Co, the crystal structure changes from the hcp to the fcc
with increasing temperature.  We present here the results for the fcc
Co.  Figure 5 shows an example of calculated DOS in the ferromagnetic 
state.  The $d$ DOS are split into up and down parts.  
The $d$ DOS in the static approximation are
smoothed and are
broadened due to thermal spin fluctuations.  The dynamical effects
reduce the exchange splitting and the band width of each spin component.
More important difference between the static and dynamical cases is 
that the quasiparticle peaks appear
at the Fermi level due to dynamical spin and charge fluctuations.
We also find a small hump at $\omega=-0.55$Ry in the up-spin DOS, 
which is caused by dynamical charge fluctuations.
%
%
\begin{figure}
\includegraphics[scale=0.85]{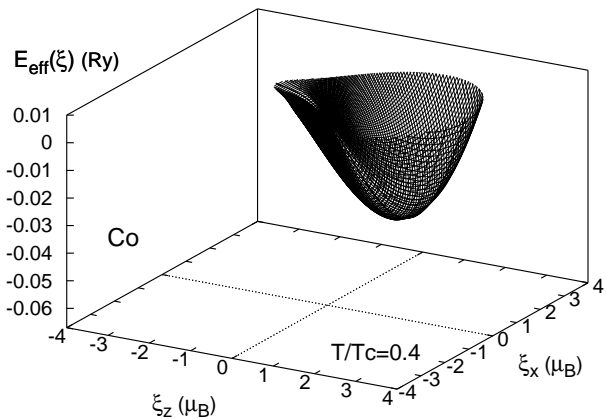}%
\caption{\label{figcomeff}
Effective potential of Co at $T/T_{\rm C}=0.4$.
}
\end{figure}
%
%

We present in Fig. 6 the effective potential for fcc Co.  The
potential has a single minimum in the ferromagnetic state.  With
increasing temperature the flat part of the potential around the origin 
extends to the negative region of $\xi_{z}$, 
and the double minimum structure appears near and above $T_{\rm C}$.  
The dynamical corrections to the effective potential
show a butterfly structure in which the ala in the region $\xi_{z} > 0$
is higher than that in $\xi_{z} < 0$, as shown in Fig. 7.
It indicates that the dynamical effects act to reduce the magnetization
and to enhance the transverse spin fluctuations.
%
%
\begin{figure}
\includegraphics[scale=0.85]{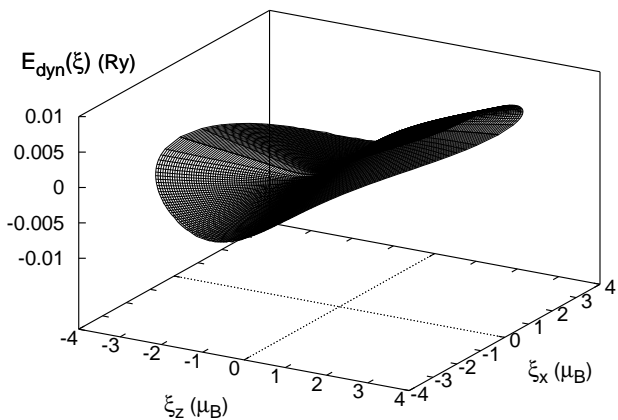}%
\caption{\label{figcoedyn}
Dynamical contribution to the effective potential in Co at 
$T/T_{\rm C}=0.4$.
}
\end{figure}
%
%

Calculated magnetization vs temperature curves for fcc Co 
are shown in Fig. 8.  We
find the Curie temperature in the static approximation 
$T_{\rm C}=3160$ K.  It is much lower than the Hartree-Fock one 
(12100 K).
The 4th-order dynamical CPA reduces $T_{\rm C}$ by 610K, and yields
$T_{\rm C}=2550$K.  The latter is overestimated by a factor of 1.8
as compared with the experimental value~\cite{colvin65} 1388K.
The inverse susceptibility follows the Curie-Weiss law though it is
considerably upward convex.
Calculated effective Bohr magneton numbers at $T/T_{\rm C} \sim 1.1$ are
obtained to be 2.4 $\mu_{\rm B}$ in the static approximation 
and 3.0 $\mu_{\rm B}$ in the 4th-order dynamical CPA.  
The latter agrees well with the experimental value 3.2 $\mu_{\rm B}$.
%
%
\begin{figure}
\includegraphics[scale=0.75]{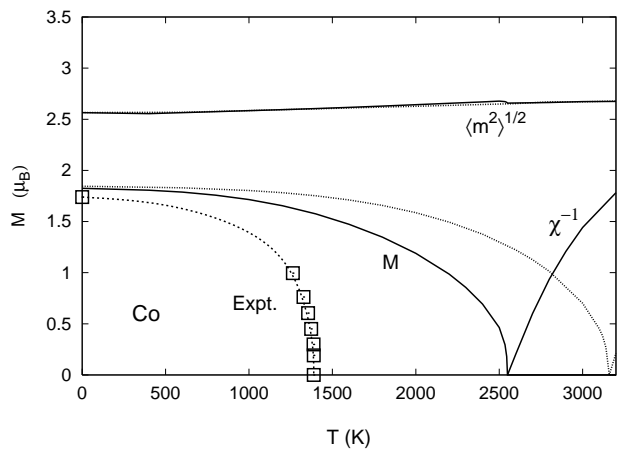}%
\caption{\label{figcomt}
Magnetization vs temperature curves ($M-T$), inverse susceptibilities 
($\chi^{-1}$), and the amplitudes of local magnetic moments 
($\langle \boldsymbol{m}^{2} \rangle^{1/2}$) for Co in the static
 approximation (dotted curves) and the 4th-order dynamical CPA 
(solid curves), respectively. Experimental data of 
magnetizations~\cite{meyers51} for 
fcc Co are shown by open squares. Dashed line shows a guide for the eye
 to an experimental $M-T$ curve. 
}
\end{figure}
%
%
The amplitude of local moment hardly change with increasing
temperatures, and has a value 2.66 $\mu_{\rm B}$ around $2600$K.  This
is close to the effective Bohr magneton number 2.55 in the local moment
model, but is smaller than the experimental value 3.2 $\mu_{\rm B}$.

Calculated DOS for the ferromagnetic Ni is shown in Fig. 9.  By
comparing the DOS with those in the static approximation, we find that
the dynamical effects reduce the exchange splitting by a factor of two
and cause a kink at the Fermi level corresponding to a quasi-particle
state.  On the other hand, a satellite which is found experimentally
around 6 eV below the Fermi 
level~\cite{himp79,east80,eber80,mar84,naka91} 
does not appear.  It indicates that
the dynamical corrections up to the 4th order with asymptotic
approximation are not enough to describe
the charge fluctuations at low temperatures $T \lesssim 500$K,
though the 6 eV satellite is found 
at $\omega = -0.45$Ry in the DOS when
calculated at high temperatures in the paramagnetic state
(see the thin dashed curve and thin solid curve in Fig. 9).
%
%
\begin{figure}
\includegraphics[scale=0.75]{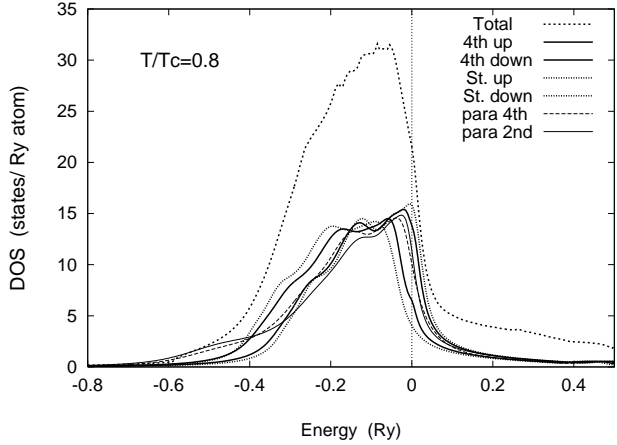}%
\caption{\label{fignidos}
Up and down $d$ partial DOS of Ni in the static approximation
 (dotted curves) and in the 4th-order dynamical CPA (solid curves).
The total DOS in the 4th-order dynamical CPA is shown
 by dashed curve. The $d$ DOS in the paramagnetic state ($T=2000$K) are
 also shown by thin dashed curve (4th-order dynamical CPA) and thin
 solid curve (2nd-order dynamical CPA).
}
\end{figure}
%
%

The effective potential in Ni shows a single minimum in both the ferro-
and the para- magnetic states as shown in Fig. 10, indicating small thermal
spin fluctuations around the equilibrium points.  
The dynamical contribution to the effective
potential in the ferromagnetic Ni shows a flat structure with a slope
along $\xi_{z}$ direction (see Fig. 11); 
$E_{\rm dyn}(\boldsymbol{\xi}) \sim \xi_{z}h_{\rm eff}$ where 
$h_{\rm eff}$ is an effective field.
This implies that the dynamical effects act as an effective magnetic
field which weakens the spin polarization.
%
%
\begin{figure}
\includegraphics[scale=0.85]{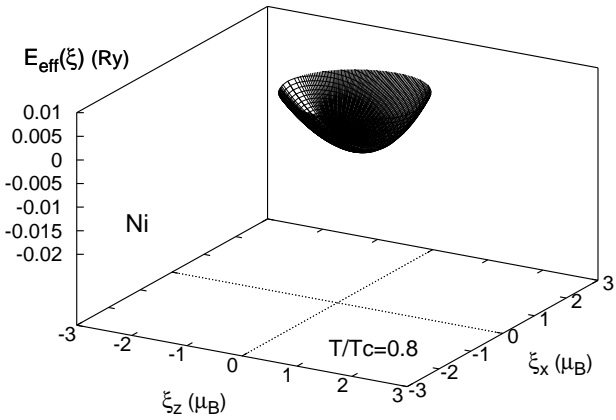}%
\caption{\label{fignieff}
Effective potential of Ni at $T/T_{\rm C}=0.8$.
}
\end{figure}
%
%
%
%
\begin{figure}
\includegraphics[scale=0.85]{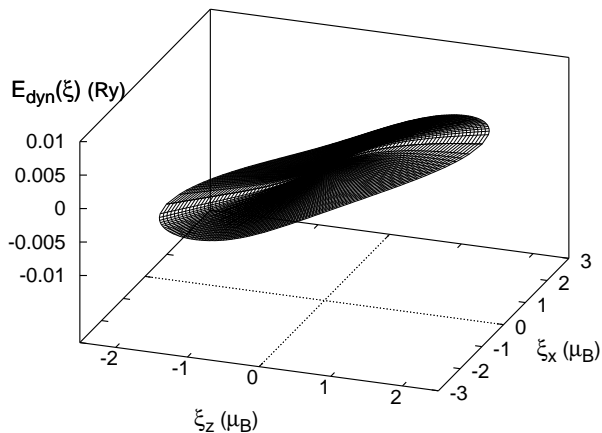}%
\caption{\label{figniedyn}
Dynamical contribution to the effective potential in Ni at 
$T/T_{\rm C}=0.8$.
}
\end{figure}
%
%

The magnetization vs temperature curves are presented in Fig. 12.  We
find calculated Curie temperatures: 1420K (static approximation),
1260K (2nd-order dynamical CPA), and 620K (4th-order dynamical CPA),
respectively.  These results are much lower than the Hartree-Fock 
value 4940 K.
The 4th-order result is in good agreement with the experimental
value~\cite{noakes66} 630K.  
The stability of the ferromagnetism in Ni is sensitive to the change 
of the DOS at the Fermi level according to the ground-state theory of 
the ferromagnetism in the low density limit~\cite{kana63}.
A large reduction of $T_{\rm C}$ due to 4th-order dynamical
corrections seems to be associated with the instability of the
ferromagnetism due to the reduction of the DOS at 
the Fermi level.  In fact,
the band width of the $d$ DOS in the 4th-order dynamical CPA is
broader than that of the the 2nd-order one, and the height of the peak in
the 4th order DOS is smaller than the 2nd-order one as shown in Fig. 9.
%
%
\begin{figure}
\includegraphics[scale=0.75]{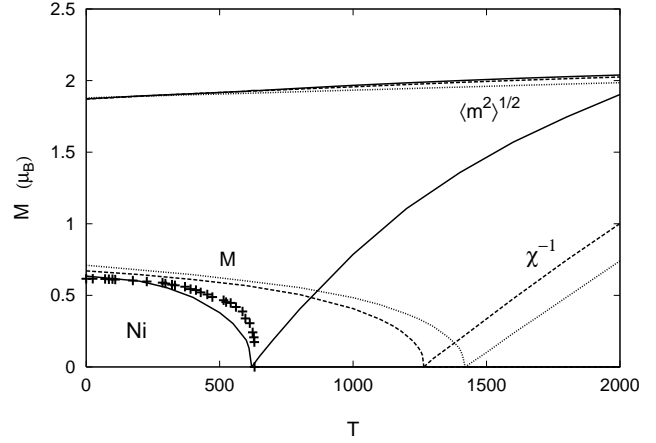}%
\caption{\label{fignimt}
Magnetization vs temperature curves ($M-T$), inverse susceptibilities 
($\chi^{-1}$), and the amplitude of local magnetic moments 
($\langle \boldsymbol{m}^{2} \rangle^{1/2}$) for Ni in the static
 approximation (dotted curves), the 2nd-order dynamical CPA 
(dashed curves), and the 4th-order dynamical CPA (solid curves), 
respectively. Experimental data of magnetization curve~\cite{weiss26} 
are shown by $+$. 
}
\end{figure}
%
%

Calculated inverse susceptibilities in Ni follow the Curie-Weiss law in the
static approximation and in the 2nd-order dynamical CPA, but show an upward
convexity in the case of the 4th-order dynamical CPA.  The upward
convexity in the high-temperature region is found in the
experimental data~\cite{fallot44}, though it is not the case in Co.  
Effective Bohr magneton numbers calculated at 
$T \sim 2000$K are 1.2 $\mu_{\rm B}$ in the static approximation as well
as in 
the 2nd-order dynamical CPA, while 1.6 $\mu_{\rm B}$ in the 4th-order 
dynamical CPA.  The latter is in good agreement with the
experimental value~\cite{fallot44} 
1.6 $\mu_{\rm B}$.  Calculated amplitude of local
moment $\langle \boldsymbol{m}^{2} \rangle^{1/2}$ slightly increases with
increasing temperature and takes a value 1.97 $\mu_{\rm B}$ at 1000K, 
which is larger than 1.27 $\mu_{\rm B}$, the value in the local moment 
model.  Because of the hybridization of the $d$ bands with the $sp$ bands,
the $d$ electron number in the metallic state ($n_{d}=8.7$) is smaller
than $n_{d}=9.4$ expected from a $d$ band model with the strong 
ferromagnetism,
so that the amplitude of the local magnetic moment is larger than the
value expected from the local moment model, 
$\sqrt{M(0)(M(0)+2)}$ where $M(0)$ denotes the ground-state magnetization.
%
%
\begin{figure}
\includegraphics[scale=0.75]{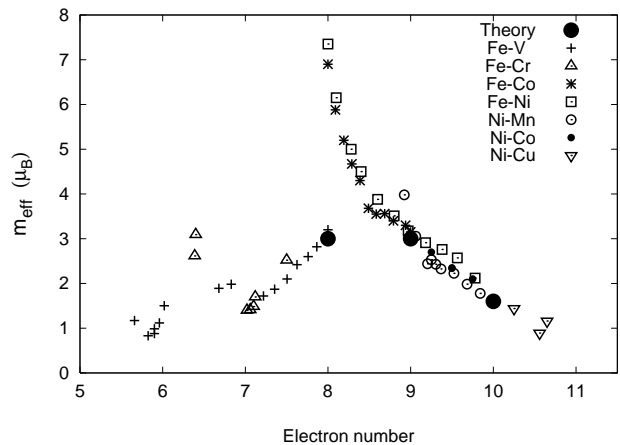}%
\caption{\label{figmeff}
Effective Bohr magneton numbers in various 3$d$ transition metal alloys 
as a function of conduction electron number per atom~\cite{wohl83,arajs62,lam63,nakagawa56,bar63,naka57,che62,kaya40,kouvel60,hicks69}.
Large closed circles show theoretical values of Fe, Co, and Ni obtained
 in the present calculations.
}
\end{figure}
%
%
\section{Summary and Discussions}
We have investigated the ferromagnetic properties of Fe, Co, and Ni on
the basis of the first-principles dynamical CPA combined with the LDA +
$U$ Hamiltonian in the TB-LMTO representation.  The dynamical CPA
takes into account single-site spin and charge fluctuations.  We adopted
the harmonic approximation to solve the impurity problem in the
self-consistent dynamical CPA.  In the harmonic approximation, 
we start from the high
temperature approximation, {\it i.e.}, the static approximation, and
take into account individually the dynamical contributions from the 
dynamical potentials with a given frequency.  We have performed the numerical
calculations taking into account the dynamical effects up to the 4th
order in Coulomb interaction.  Calculated effective potential
shows up the double minimum structure in case of Fe, the single minimum
structure in case of Ni, and the Co is in between the two.

We found that dynamical effects reduce the exchange splitting in the
ferromagnetic DOS, suppress the band broadenings due to thermal
spin fluctuations in the static approximation, and cause the `6 eV'
satellites.  In the case of Co and Ni, we also found the quasiparticle
peaks at the Fermi level.  

We calculated the magnetization vs temperature curves.  Curie
temperatures obtained from the 4th-order dynamical CPA 
are 1930K for Fe, 2550K for fcc Co, and 620K for Ni.
Although calculated $T_{\rm C}$ in Ni is close to the experimental value
630K, those in Fe and Co are overestimated by a factor of 1.8 
in comparison with the experimental values, 
1040K (Fe) and 1388K (Co), respectively.
In the present calculations, the ground state magnetizations obtained by
an extrapolation are 2.58 $\mu_{\rm B}$ (Fe), 1.82 $\mu_{\rm B}$ (Co),
and 0.63 $\mu_{\rm B}$ (Ni), which are considerably larger than the
experimental values, 2.22 $\mu_{\rm B}$ (Fe)~\cite{danan68}, 1.74 
$\mu_{\rm B}$ (Co)~\cite{bes70},
and 0.62 $\mu_{\rm B}$ (Ni)~\cite{danan68}.  
These facts suggest that the present
calculations tend to underestimate the screening of Coulomb
interactions, especially at low temperatures.  
It is desirable to take into account dynamical effects more
accurately in order to suppress the magnetization below $T_{\rm C}$.  
One has also to take into
account inter-site spin correlations in order to realize quantitative
description of the Curie temperature, 
going beyond the single-site approximation.

We have calculated the paramagnetic susceptibilities.  Calculated
effective Bohr magneton numbers, 3.0 $\mu_{\rm B}$ (Fe), 3.0
$\mu_{\rm B}$ (Co), and 1.6 $\mu_{\rm B}$ (Ni),
quantitatively explain the experimental data as shown in
Fig. 13.  Experimental data of effective Bohr magneton numbers in 3$d$
transition metal
alloys~\cite{wohl83,arajs62,lam63,nakagawa56,bar63,naka57,che62,kaya40,kouvel60,hicks69} 
continuously distribute including the three points for
Fe, Co, and Ni.  It is a future problem to investigate whether or not
the dynamical CPA can quantitatively describe these data after
extension of the theory to alloys.

\section*{Acknowledgment}
The present work is supported by Grant-in-Aid for Scientific Research
(22540395). 
Numerical calculations have been partly carried out with use of the
SGI Altix ICE 8400EX in the Supercomputer Center, Institute of Solid State
Physics, University of Tokyo.

\end{document}